\documentclass[aps,prl,reprint,showpacs,twocolumn,amsmath,amssymb,10pt,floatfix]{revtex4-1}
\usepackage[utf8]{inputenc}
\usepackage{color,graphicx}
\usepackage{physics,bm}
\usepackage{hyperref}
\usepackage{dsfont}

\newcommand{\rref}[1]{Ref.~\cite{#1}}
\newcommand{\rrefs}[1]{Refs.~\cite{#1}}
\newcommand{\fref}[1]{Fig.~\ref{#1}}
\newcommand{\eref}[1]{Eq.~(\ref{#1})}
\newcommand{\bdyad}[3]{\left[\dyad{#1}{#2}\right]_{#3}}

\begin{document}
\title{Reservoir engineering with ultracold Rydberg atoms}

\author{David~W.~Sch{\"o}nleber}
\author{Christopher~D.~B.~Bentley}
\author{Alexander~Eisfeld}
\affiliation{Max Planck Institute for the Physics of Complex Systems, N\"othnitzer Strasse 38, 01187 Dresden, Germany}

\begin{abstract}
We apply reservoir engineering to construct a thermal environment with controllable temperature in an ultracold atomic Rydberg system.
A Boltzmann distribution of the system's eigenstates is produced by optically driving a small environment of ultracold atoms, which is coupled to a photonic continuum through spontaneous emission.
This technique provides a useful tool for quantum simulation of dynamics coupled to a thermal environment.
Additionally, we demonstrate that pure eigenstates, such as Bell states, can be prepared in the Rydberg atomic system using this method.
\end{abstract}

\maketitle

\emph{Introduction.}---Solving the dynamics of open quantum systems is one important task addressed by quantum simulators~\cite{feynman1982,Lloy96S,buluta2009,georgescu2014}.
The dynamics generated by the environment-couplings of the target system must be reproduced in the simulator~\cite{Lloy96S,Tsen00PRA}.
Coherent~\cite{Lloy01PRA,Klie11PRL} and incoherent approaches~\cite{Baco01PRA} have been developed for this \emph{reservoir engineering} problem, for particular environments or degrees of freedom in the system.
Reservoir engineering~\cite{Poya96PRL} involves design of the environment or its coupling to the system, which can be applied in a variety of systems for entanglement generation and protection~\cite{Beig00JMO,Carv01PRL,kraus2008,Carv08PRA,Krau11PRL,Cho11PRL,Stan12NJP,Lin13N,Shan13N,Reit13PRA,bentley2014,morigi2015}, dissipative computation~\cite{verstraete2009} and open quantum system simulation~\cite{diehl2008,weimer2010,barreiro2011,Schi13NP,Piil06PRA,mostame2012}.   
Here, we consider reservoir engineering of an extreme nature: using optical control, we transform a typically non-thermal environment into a thermal environment with controllable temperature, such that the system relaxes to a corresponding mixture of its eigenstates.

\begin{figure}[tb]
\centering
\includegraphics[width=\columnwidth]{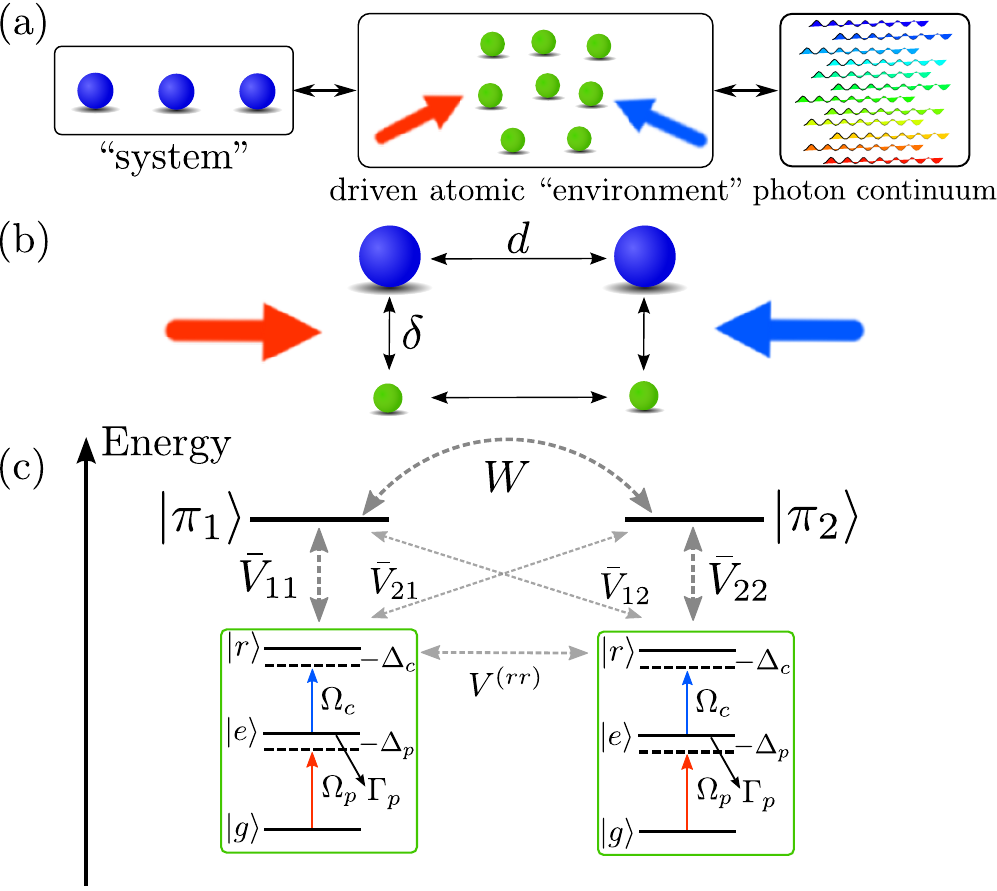}
\caption{(Color online) (a) Conceptual sketch of the setup. A ``system" of Rydberg atoms is coupled to a laser-driven atomic ``environment'', which is dissipative through its coupling to the electromagnetic continuum. (b) Possible implementation for the case of two system atoms. Laser-driven environment atoms are placed at a distance $\delta$ from each of the system atoms, which have interatomic separation $d$. (c) Level sketch. The dimer states $\ket{\pi_1}$ and $\ket{\pi_2}$ (see main text) are coupled to each other via resonant dipole-dipole interaction with strength $W$ and interact with the Rydberg states $\ket{r}$ of the environment atoms via the interactions $\bar{V}_{n\alpha}$. The environment atoms are laser-driven, realizing the level scheme depicted in the green boxes.}
\label{fig:sketch_setup}
\end{figure}

In this Letter we show that a driven dissipative atomic environment, as shown in Figure~\ref{fig:sketch_setup}(a), provides a highly-tunable environment.
In our setup, the steady-state populations of the system eigenstates can be precisely controlled by means of currently-achievable frequencies and intensities of the two lasers driving the environment atoms.
We prepare system eigenstates and thermal states (i) on a timescale shorter than the system decay timescale and (ii) in a long time limit. 
While (i) demonstrates dissipative state preparation of entangled and thermal states, (ii) can be used to mimic a thermal environment for a quantum simulator.
We highlight that the effective temperature scale of the Boltzmann distribution of eigenstates is determined by the system interaction strength rather than the `ambient' temperature of the ultracold environment.

\emph{Setup.}---We consider the setup sketched in \fref{fig:sketch_setup}(a) and described in detail in the Supplemental Material (SM)~\cite{suppinf} and \rref{schoenleber2015}. Dipole-coupled Rydberg atoms constitute our system of interest. 
Due to their strong interactions and relative ease to laser-control and position experimentally, Rydberg systems have been proposed for quantum simulation of spin Hamiltonians~\cite{weimer2010,lesanovsky2012}, electron-phonon coupling~\cite{hague2012}, as well as exciton transport~\cite{muelken2007,guenter2013,barredo2015,schoenleber2015,schempp2015,genkin2016}. 
The environment for the Rydberg system is provided by laser-driven atoms, which in turn are coupled to a continuum of electromagnetic modes, thereby inducing radiative transitions to lower-energy states (spontaneous emission) in the atomic environment. Interactions between Rydberg system and environment atoms are introduced via Rydberg states of the environment atoms. The tunability of the environment arises through its composition of a finite part, the laser-driven three-level atoms, and an infinite part, the photon bath. By tuning the parameters of the lasers addressing the environment atoms, the dynamics within the finite part, and in particular the timescale of this dynamics, can be controlled.

We now exemplify the relevant properties of our setup for a Rydberg dimer, i.e., two Rydberg atoms as shown in \fref{fig:sketch_setup}(b). The atoms are prepared in two different Rydberg states. Here we consider one atom prepared in a Rydberg $\ket{s}=\ket{\nu s}$ state with principal quantum number $\nu$ and angular momentum quantum number $\ell=0$, and the other atom prepared in a Rydberg $\ket{p}=\ket{\nu p}$ state with angular momentum $\ell=1$. 
Resonant dipole-dipole interactions of strength $W$ lead to excitation migration \cite{robicheaux2004,barredo2015}, described by the Hamiltonian
\begin{equation}\label{eq:agg_H_dimer}
 \mathcal{H}_\mathrm{sys} = W(\dyad{ps}{sp} + \mathrm{H.c.}) \equiv W(\dyad{\pi_1}{\pi_2} + \mathrm{H.c.}),
\end{equation}
where H.c. denotes the Hermitian conjugate and we introduce the notation $\ket{\pi_1}=\ket{ps}$ ($\ket{\pi_2}=\ket{sp}$) to denote the location of the $\ket{p}$ excitation within the system.
We will later be concerned with the eigenvalues of this Hamiltonian, given by $E_{\pm} = \pm W$, and the corresponding Bell eigenstates $\ket{\Psi^\pm} = (\ket{\pi_1}\pm\ket{\pi_2})/\sqrt{2}$.  
For simplicity, we do not include the finite lifetime of the system in our calculations.

The environment for the Rydberg system is provided by laser-driven atoms. For simplicity, we place each at a distance $\delta$ from a given system-atom, such that the vectors along $d$ and $\delta$ respectively enclose a right angle. The environment atoms are addressed by two laser beams. The first laser, with Rabi frequency $\Omega_p$ and detuning $\Delta_p$, couples the ground state $\ket{g}$ of an environment atom to a short-lived intermediate state $\ket{e}$ with radiative decay rate $\Gamma_p$. The second laser, with Rabi frequency $\Omega_c$ and detuning $\Delta_c$, couples the intermediate state $\ket{e}$ to a Rydberg state $\ket{r}\neq\ket{p},\ket{s}$.

The Rydberg states $\ket{r}$ of the environment atoms introduce interactions both between the environment atoms and between the environment atoms and system atoms. The environment atoms interact with each other via van der Waals interaction $V^{(rr)}$. The interactions between environment and system atoms are state-dependent, 
\begin{equation}
 \mathcal{H}_\mathrm{int} = \sum_{n,\alpha}{\bar{V}_{n\alpha} \dyad{\pi_n}{\pi_n}} \bdyad{r}{r}{\alpha},
\end{equation}
where $\bar{V}_{n\alpha} = V^{(pr)}_{n\alpha} + \sum_{m\neq n} V^{(sr)}_{m\alpha}$ denotes the overall interaction of a specific environment atom $\alpha$ with the entire system if the latter is in the state $\ket{\pi_n}$. Here, $V^{(pr)}_{n\alpha}$ ($V^{(sr)}_{n\alpha}$) indicates the interaction between the $\ket{r}$ state of environment atom $\alpha$ with a $\ket{p}$ ($\ket{s}$) excitation at system atom $n$. Due to their distance-dependence, the interactions $V^{(pr)}_{n\alpha}$ and $V^{(sr)}_{n\alpha}$ increase drastically with decreasing distance and are therefore strongest for adjacent environment and system atoms, i.e., $\alpha=n$.

The resulting level scheme for a Rydberg dimer is shown in \fref{fig:sketch_setup}(c). 
For state engineering to be feasible, it is essential that the interaction between an environment Rydberg state $\ket{r}$ and the states $\ket{p}$ and $\ket{s}$ of the adjacent system atom are different, e.g.\ obeying $|V_{nn}^{(pr)}|\gg |V_{nn}^{(sr)}|$. The numerical values of the interactions are the same as in \rref{schoenleber2015} and are detailed in the SM~\cite{suppinf}. In the following, we choose the environment-atom geometry such that $d=5~\mu$m and $\delta=2~\mu$m. 

\emph{Bell state preparation.}---We now show that the Rydberg dimer can be dissipatively prepared in the Bell eigenstates $\ket{\Psi^\pm}$ of the system Hamiltonian, \eref{eq:agg_H_dimer}. To this end, we numerically optimize the laser parameters $(\Omega_p,\Delta_p,\Omega_c,\Delta_c)$ for the fixed geometry. Note that the optimized laser parameters are independent of the initial state of system or environment. We thus use a system (environment) state that is easy to access experimentally in our numerical simulations, given by $\ket{\pi_1}$ ($\ket{gg}$).

\begin{figure}[tb]
\centering
\includegraphics[width=\columnwidth]{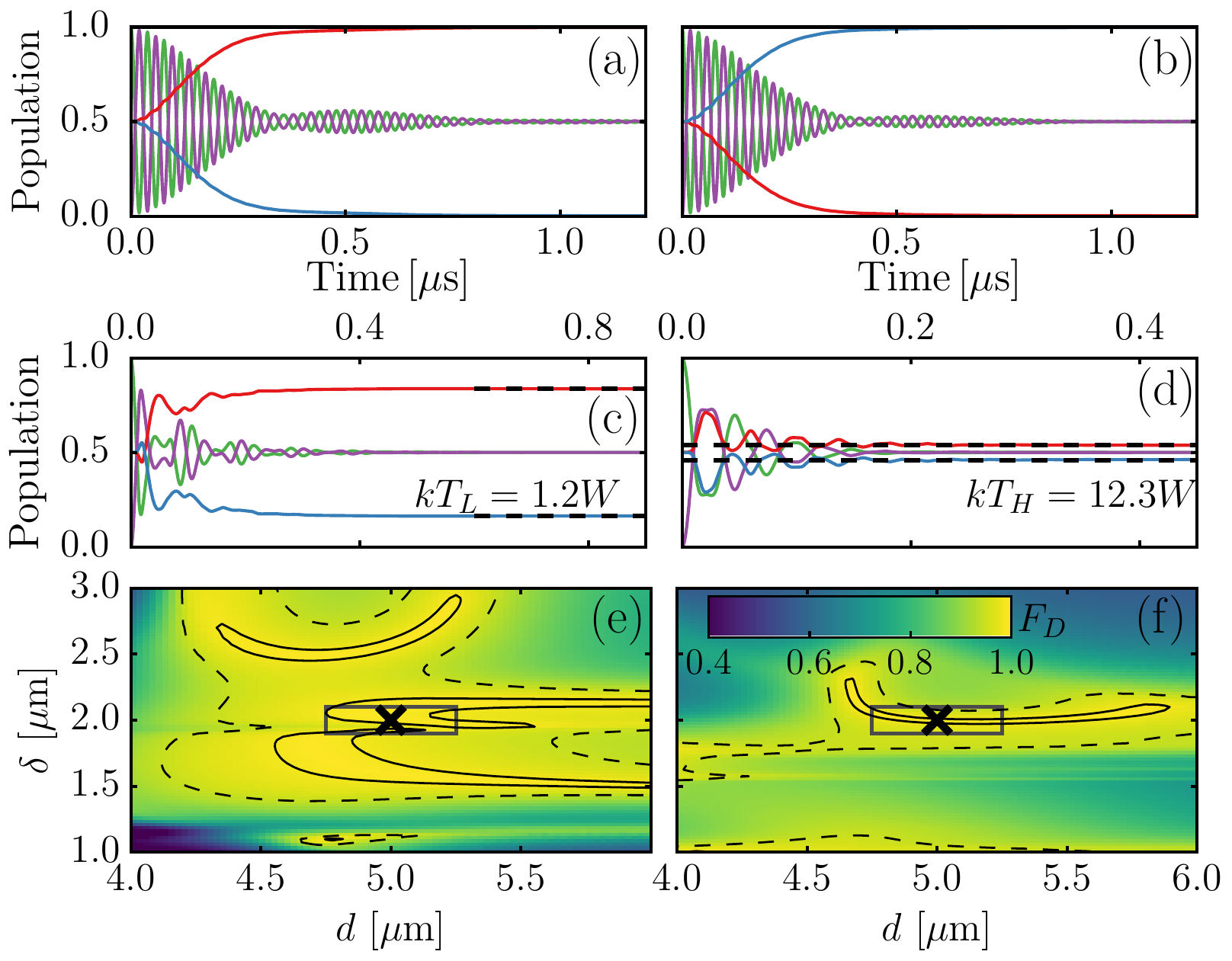}
\caption{(Color online) (a,b) Bell state preparation in a Rydberg dimer. Population dynamics of anti-symmetric (a) and symmetric (b) Bell state. Red (blue) lines indicate the population of the anti-symmetric (symmetric) Bell state and green (purple) lines indicate the populations of the states $\ket{\pi_1}$ ($\ket{\pi_2}$).
(c-f) Thermal state preparation in a Rydberg dimer with temperatures $T_L$ (c,e) and $T_H$ (d,f). (c,d) Population dynamics. Solid red (blue) lines denote the population of the lower (higher) energy eigenstate while green (purple) lines indicate localized state $\ket{\pi_1}$ ($\ket{\pi_2}$) population. The target populations are shown with black dashed lines. (e,f) $F_D$ for thermal state preparation as a function of the dimer separation $d$ and the dimer-environment atom distance $\delta$. Crosses mark the values used in the upper plots. The gray boxes indicate a $\pm 5\%$ uncertainty in the distances. The solid (dashed) lines mark the contours defined by $F_D=0.99$ ($F_D=0.95$). Laser parameters are listed in Tab.~I in the SM.}
\label{fig:full_preparation}
\end{figure}

In \fref{fig:full_preparation} we illustrate the preparation of the anti-symmetric (a) as well as the symmetric (b) Bell state. The marked revival feature in the population dynamics is absent in the dynamics of the eigenstate populations, indicating that the eigenstates can indeed be selectively addressed by the dissipative environment. 

To assess the difference between the target density matrix, in this case the projector on the appropriate Bell state, and the density matrix obtained after numerical propagation, we employ two often-adopted distance measures (cf.\ \rrefs{nielsen2010,gilchrist2005}). The first measure is the \emph{fidelity} $F(\rho_1,\rho_2)$, and the second measure $F_D(\rho_1,\rho_2)$ is related to \emph{trace distance} $D(\rho_1,\rho_2)$ via $F_D(\rho_1,\rho_2) = 1-D(\rho_1,\rho_2)$. Details are given in the SM~\cite{suppinf}. To illustrate the different timescales on which steady-state preparation is feasible, we evaluate the two distance measures $F$ and $F_D$ at $t=1~\mu$s and $t\rightarrow \infty$ respectively, denoting the $t\rightarrow \infty$ case by $\tilde{F}$ and $\tilde{F}_D$.
Table~I in the SM shows that high-fidelity preparation can be achieved on a timescale of about $1~\mu$s, which is much shorter than the lifetime of the chosen dimer states, which is approximately 56~$\mu$s~\cite{beterov2009}.

Note that we prepare Bell states in the Rydberg manifold, as opposed to other proposals~\cite{browaeys2016,saffman2016}, which involve a ground state contribution. 

\emph{Thermal state preparation.}---In addition to the preparation of a single system eigenstate, we can also prepare mixtures of eigenstates. To illustrate this, we consider thermal states \cite{yung2010}
\begin{equation}
 \rho^\mathrm{th}_T = \frac{1}{Z}\sum_n{e^{-\mathcal{H}_\mathrm{sys}/(kT)} \dyad{\varphi_n}},
\end{equation}
where $Z = \Tr\{e^{- \mathcal{H}_\mathrm{sys}/(kT)} \dyad{\varphi_n}\}$, $\ket{\varphi_n}$ denotes the eigenstates of $\mathcal{H}_\mathrm{sys}$, $k$ the Boltzmann constant and $T$ the temperature of the system. We emphasize that $T$ is not the ambient temperature of the environment atoms, which is typically $\sim \mu$K. Note that the relevant energy scale providing the temperature scale is the resonant dipole-dipole interaction $W$. 
For a dimer, the two system eigenstates are separated in energy by $2W$, such that $\rho^\mathrm{th}_T \propto (\dyad{\varphi_1} + e^{-2W/(kT)}\dyad{\varphi_2})$, where $\ket{\varphi_1}$ and $\ket{\varphi_2}$ are the dimer eigenstates $\ket{\Psi^-}$ and $\ket{\Psi^+}$, respectively.
To illustrate the versatility of the scheme we demonstrate low and high temperature mixtures, given by $kT_L = 1.2~W$ and $kT_H = 12.3~W$ respectively. The numerical values are chosen such that $T_L$ yields strong asymmetries in the eigenstate populations while $T_H$ yields similar occupations of the eigenstates.

For a Rydberg dimer, the preparation of thermal states with temperature $T_L$ (c) and $T_H$ (d) is shown in \fref{fig:full_preparation}, using exemplary laser parameter sets. Both states can be prepared with high fidelity (cf.\ Tab.~I in the SM).

\emph{Robustness.}---For the target state to be experimentally accessible, it is important that small variations in the laser parameters or distances do not lead to a strong reduction of the fidelity. We have verified this by varying both distances (dimer separation $d$ and dimer-environment atom distance $\delta$) as well as laser parameters. Generally, the laser and distance parameter values required to obtain high target-state fidelities are non-unique, reflecting the grand tunability of our environment. 
In particular, there are possible parameter variations around a given parameter set for which the fidelity does not drop significantly. The parameter range for the respective variations depends on the desired target state. To illustrate this, we show in \fref{fig:full_preparation}(e) and (f) the dependence of $F_D$ of the two thermal dimer states displayed in \fref{fig:full_preparation}(c) and (d) on the distances $d$ and $\delta$.
Although the high-fidelity parameter space differs significantly for the two different thermal dimer states, in both cases small variations in the order of a few percent in the distances $d$ and $\delta$ still allow for high-fidelity preparation after a preparation time of $t=1~\mu$s.
The dynamics towards thermal equilibrium depends on the chosen laser parameters, and is different in general for different laser parameter sets.

\begin{figure}[tb]
\centering
\includegraphics[width=\columnwidth]{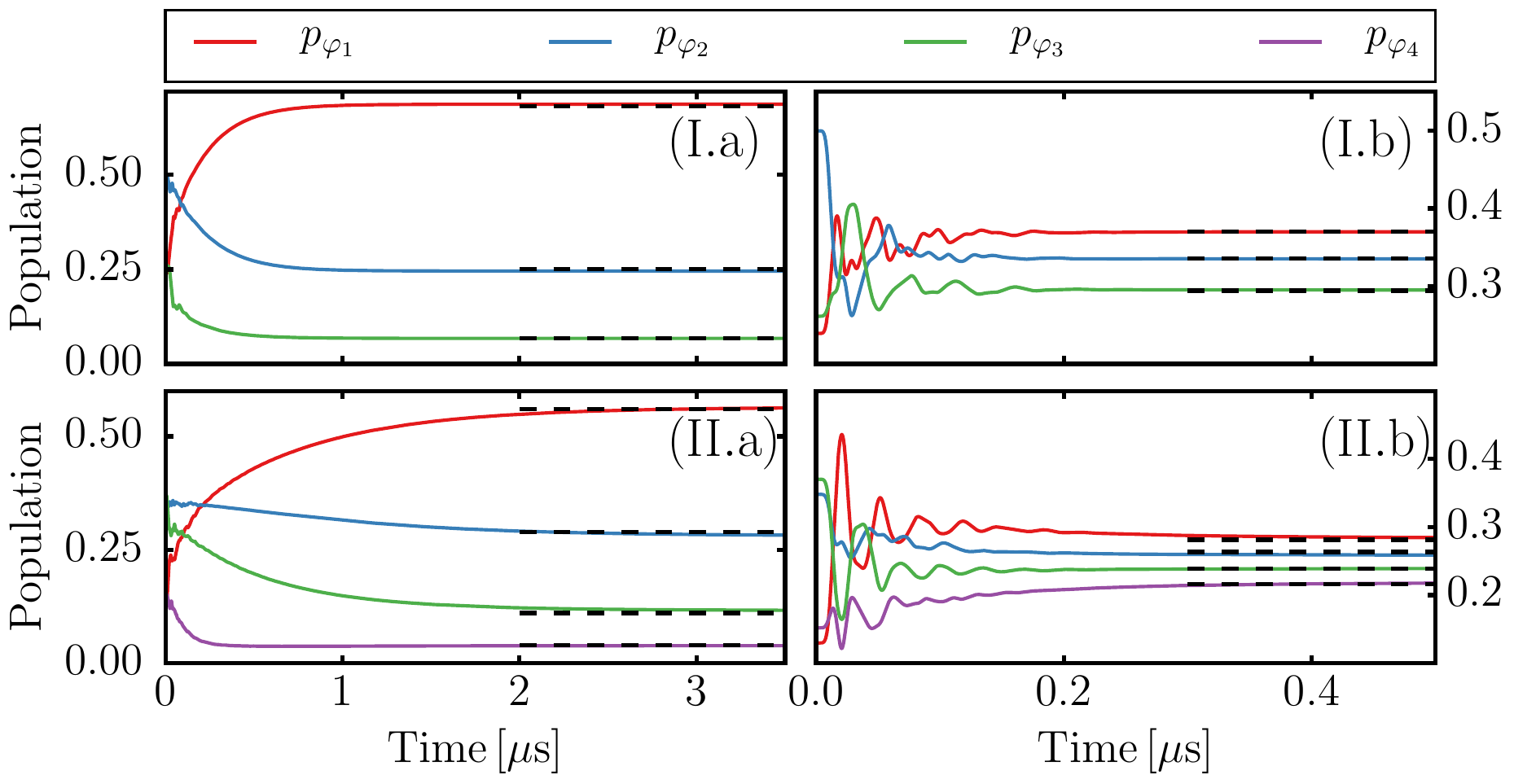}
\caption{(Color online) Thermal state preparation in a trimer (I) and tetramer (II) with temperatures $T_L$ (a) and $T_H$ (b). Solid colored lines show the populations $p_{\varphi_n}$ of the system eigenstates $\ket{\varphi_n}$, and the target populations are indicated with dashed black lines. Laser parameters are listed in Tab.~I in the SM.}
\label{fig:thermal_trimer_tetramer_preparation}
\end{figure}

\emph{Results for larger systems.}---Dissipative state-preparation is not limited to a Rydberg dimer. To demonstrate this, we extend the setup shown in \fref{fig:sketch_setup}(b) by placing $N\!>\!2$ system atoms on an equidistant, one-dimensional lattice (lattice constant $d$). The environment atoms are placed on a copy of this lattice, shifted by the distance $\delta$ orthogonal to the system lattice. In \fref{fig:thermal_trimer_tetramer_preparation} accordingly we show the preparation of thermal states with temperatures $T_L$ and $T_H$ for a Rydberg trimer ($N\!=\!3$ system atoms) as well as a tetramer ($N\!=\!4$ system atoms) using exemplary laser parameter sets. In both cases, the thermal states can be prepared with high fidelity (cf.\ Tab.~I in the SM).
To verify further scalability of our approach, we have evaluated $F_D$ after $1~\mu$s and $2~\mu$s for thermal-state preparation with temperatures $T_H$ and $T_L$, respectively,  for up to 6 system and environment atoms (cf.\ Fig.~2 in the SM), using the optimized parameters for $N\!=\!4$. We find a weak dependence of $F_D$ on system size, indicating scalable state preparation. 
Note, however, that state-preparation at short timescales becomes more difficult for larger system sizes. In particular, for constant atom spacing, the energy differences between the system eigenenergies decrease with increasing system size as $1/N^2$. Larger timescales are thus required for distinctive dynamics for individual eigenstates to emerge as needed for the preparation of low-temperature thermal states.

\emph{Mechanism.}---The underlying mechanism for preparation of the Bell and thermal steady states can be traced back to the appearance of quasi-resonances.
This is illustrated most clearly for the case of Bell state preparation:
to prepare a specific Bell state, we want to ensure that (i) there is negligible population transfer out of this target state, while (ii) population is transfered from the non-target Bell state to the target state. 

\begin{figure}[tb]
\centering
\includegraphics[width=\columnwidth]{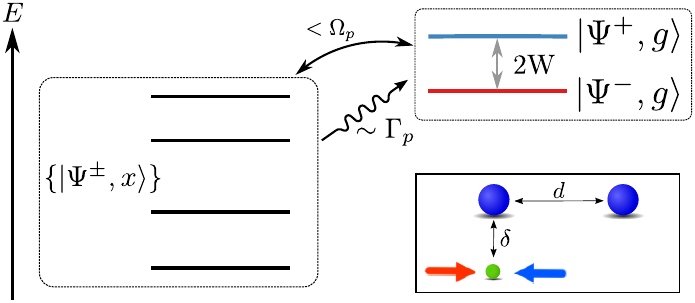}
\caption{(Color online)  Sketch of the energy levels (not to scale) of the combined basis $\{\ket*{\Psi^\pm, g},\ket*{\Psi^\pm,x}\}$ with $x\in\{e,r\}$ in a configuration that prepares the $\ket*{\Psi^{+}}$ state. For this preparation, the highest-energy state of the $\{\ket{\Psi^\pm,x}\}$ manifold is quasi-resonant with the non-target state $\ket*{\Psi^-,g}$. The wiggly line indicates that spontaneous emission feeds population from the $\{\ket{\Psi^\pm,x}\}$ manifold to the states with a ground state contribution while the curved arrow indicates coherent coupling between the states of the $\{\ket{\Psi^\pm,g}\}$ and the $\{\ket{\Psi^\pm,x}\}$ manifolds. More details are provided in the SM. The inset shows a sketch of the corresponding setup.}
\label{fig:sketch_mechanism}
\end{figure}

In the following we discuss exemplarily for the preparation of the $\ket{\Psi^+}$ state how the above requirements (i-ii) can be fulfilled. To simplify the discussion we consider the smallest system that allows dissipative state preparation, i.e, a Rydberg dimer with a single environment atom (see inset in \fref{fig:sketch_mechanism}).

For the preparation of the $\ket{\Psi^+}$ state, a convenient choice of the target system-environment state is the product state $\ket{\Psi^+,g}$. This state, which contains the desired system state, does not lose population due to spontaneous emission. 
To guarantee that our target state will emerge as a steady state in the long-time dynamics, we further ensure that negligible population transfer out of this state occurs in the coherent evolution. This can be done by energetically detuning our target state $\ket{\Psi^+,g}$ from all other states to which it couples, thereby satisfying condition (i).
The second condition (ii) can be met by engineering a quasi-resonance between the non-target Bell state $\ket{\Psi^-,g}$ and a state of the $\{\ket{\Psi^\pm,x}\}$ manifold, with $x \in \{e,r\}$. That way, population is transferred from the non-target state to the $\{\ket{\Psi^\pm,x}\}$ manifold, from where it can decay into the $\ket{\Psi^\pm,g}$ states. This engineering can be done by choosing appropriate laser parameters; in particular an appropriate detuning $\Delta_p$.
Details on how the appropriate parameters can be obtained analytically are provided in the SM~\cite{suppinf}. There, it is also demonstrated that this choice of parameters leads to high state-preparation fidelities.

In \fref{fig:sketch_mechanism} we present a sketch of the energy levels of the states involved in the preparation of the $\ket{\Psi^+}$ state described above, illustrating both the detuning of the target $\ket{\Psi^+,g}$ state from the coherent dynamics and the quasi-resonance between non-target $\ket{\Psi^-,g}$ state and the $\{\ket{\Psi^\pm,x}\}$ manifold. 
We note that state preparation can also be achieved in other ways than the one discussed here.

For thermal state preparation, complete isolation of one target state is no longer desirable.
Since the target state is now a mixture of the system eigenstates, we prepare couplings of varying strength from both $\ket{\Psi^{\pm}, g}$ states to the subspace spanned by the states $\{\ket{\Psi^\pm,x}\}$.
Perhaps the simplest case involves coupling a single state in this subspace to both $\ket{\Psi^\pm, g}$ states. Then the relative energies of the states can be controlled to provide the requisite thermal populations.

For more system atoms, there are a larger number of states, and necessarily more complex contributions due to more couplings, which complicate analytical considerations.
Nevertheless, this quasi-resonance picture can still be used to describe the involved processes.
The precise parameter contributions in this case we determine via numerical optimization, to
provide the desired balance of population gain/loss of the system eigenstates for state preparation even for larger systems, as we have demonstrated.

\emph{Conclusions and outlook.}---We have shown that just a few laser-driven atoms already realize a tunable environment enabling tailored dissipative state preparation in a Rydberg system. In particular, thermal states with temperatures vastly different from the temperature of the ultracold atomic environment can be prepared with high fidelity, using only the laser frequencies and intensities driving the environment atoms as control parameters. 
We highlight the flexibility provided by an intermediate, controllable environment for engineering target system-environment interactions; this flexible reservoir engineering is highly desirable for quantum simulation schemes.
Detection of the spontaneous decay events in the environment atoms, which can herald state preparation, can give further flexibility to our state-preparation scheme, and can provide conditional enhancement of the state fidelity~\cite{bentley2014}.

We note that scalability of the Rydberg system to larger sizes is limited by the radiative lifetimes of the Rydberg $\ket{p}$ and $\ket{s}$ states. An experimental realization in this case would require conditioning on a non-decayed system. Using higher-lying Rydberg states with longer lifetimes could diminish the demands on the experiment.

\section{Supporting information}

\emph{System and environment Hamiltonian.}---
We consider $N$ ultracold Rydberg atoms, of which $N-1$ atoms are initially prepared in a Rydberg $\ket{s}=\ket{\nu s}$ state with principal quantum number $\nu$ and angular momentum quantum number $\ell=0$, and the remaining atom is prepared in a $\ket{p}=\ket{\nu p}$ state with $\ell=1$~\cite{schoenleber2015}. The Rydberg states $\ket{s}$ and $\ket{p}$ have a large transition dipole moment, resulting in an interaction $W_{nm}=C_3/(\bm{R}_n-\bm{R}_m)^3$ between states with a localized $\ket{p}$ excitation at atoms $n$ and $m$ with positions $\bm{R}_n$ and $\bm{R}_m$, respectively~\cite{robicheaux2004,barredo2015}. 
Ignoring van der Waals corrections~\cite{zoubi2014}, the Hamiltonian describing the dynamics within the Rydberg system is thus given by
\begin{equation}\label{eq:aggregate_Hamiltonian}
  \mathcal{H}_\mathrm{sys} = \sum_{n\neq m} W_{nm}\dyad{\pi_n}{\pi_m},
\end{equation}
where we introduce the states $\ket{\pi_n}$ with the $\ket{p}$ excitation localized at atom $n$, e.g.\ $\ket{\pi_1}=\ket{pss\cdots}$.

The environment for the Rydberg system is provided by laser-driven atoms, each placed at a distance $\delta$ from a given system atom [cf.\ Fig.~1(a) in the main text for the specific arrangement considered in the Letter]. The environment atoms are addressed by two laser beams. The first laser with Rabi frequency $\Omega_p$ and detuning $\Delta_p$ couples the ground state $\ket{g}$ of an environment atom to a short-lived intermediate state $\ket{e}$. The second laser with Rabi frequency $\Omega_c$ and detuning $\Delta_c$ couples the intermediate state $\ket{e}$ to a Rydberg state $\ket{r}\neq\ket{p},\ket{s}$. In the rotating wave approximation, the laser Hamiltonian accordingly reads as
\begin{align}\label{eq:laser_Hamiltonian}
 \mathcal{H}_\mathrm{laser} &= \sum_\alpha \left[\frac{\Omega_p}{2}\bdyad{e}{g}{\alpha} + \frac{\Omega_c}{2}\bdyad{r}{e}{\alpha} + \mathrm{H.c.}\right] \nonumber\\
 & - \Delta_p\bdyad{e}{e}{\alpha} - (\Delta_p+\Delta_c)\bdyad{r}{r}{\alpha}.
\end{align}
The environment atoms also interact among themselves via Rydberg-Rydberg interaction $V_{\alpha\beta}^{(rr)}$, described by the Hamiltonian
\begin{equation}
  \mathcal{H}_\mathrm{int,env} = \sum_{\alpha<\beta} V_{\alpha\beta}^{(rr)} \bdyad{r}{r}{\alpha}\bdyad{r}{r}{\beta},
\end{equation}
where $\alpha,\beta$ label the environment atoms. The total Hamiltonian of the environment is thus $\mathcal{H}_\mathrm{env} = \mathcal{H}_\mathrm{laser}+\mathcal{H}_\mathrm{int,env}$.

The interactions between environment and system atoms are state-dependent,
\begin{equation}
 \mathcal{H}_\mathrm{int} = \sum_{n,\alpha}{\bar{V}_{n\alpha} \dyad{\pi_n}{\pi_n}} \bdyad{r}{r}{\alpha},
\end{equation}
where $\bar{V}_{n\alpha} = V^{(pr)}_{n\alpha}+ \sum_{m\neq n} V^{(sr)}_{m\alpha}$ denotes the overall interaction of the specific environment atom $\alpha$ with the entire system if the latter is in the state $\ket{\pi_n}$. Note that Latin indices such as $n,m$ refer to system atoms while Greek indices such as $\alpha,\beta$ to environment atoms. We emphasize that for state engineering, it is essential that the interaction between the $\ket{r}$ state of an environment atom and the $\ket{s}$ and $\ket{p}$ states of an adjacent system atom differ strongly, e.g.\ obeying $|V_{nn}^{(pr)}|\gg |V_{nn}^{(sr)}|$. 

The resulting level scheme for a Rydberg dimer is shown in Fig.~1(c) in the main text, where arrows of different thickness indicate that the dominant interactions between environment atoms and system are \emph{local}, i.e., strongest between the $\ket{p}$ excitation at atom $n$ and the adjacent environment atom $n$ (at a distance $\delta$).

Including spontaneous emission with rate $\Gamma_p$ from the intermediate state $\ket{e}$ via the Lindblad super-operator
\begin{equation}
 \mathcal{L}[\rho(t)] = \sum_\alpha\frac{1}{2}\left\{[L_\alpha \rho(t),L_\alpha^\dagger] + [L_\alpha,\rho(t)L_\alpha^\dagger] \right\}
\end{equation}
with Lindblad operator $L_\alpha = \sqrt{\Gamma_p}\bdyad{g}{e}{\alpha}$, the master equation for the full density matrix $\rho$ of the system and environment takes the form ($\hbar=1$)
\begin{equation}
 \dot{\rho}(t) = -i[\mathcal{H}_\mathrm{tot},\rho(t)] + \mathcal{L}[\rho(t)], \label{eq:me}
\end{equation}
where $\mathcal{H}_\mathrm{tot}$ denotes the total Hamiltonian of system and environment, $\mathcal{H}_\mathrm{tot}=\mathcal{H}_\mathrm{sys} + \mathcal{H}_\mathrm{env} + \mathcal{H}_\mathrm{int}$.

We now specify our state choice and interactions as in \rref{schoenleber2015}.  We take the Rydberg states of the system to be $\ket{p}=\ket{43p}$ and $\ket{s}=\ket{43s}$ of $^{87}$Rb, for which $C_3/(2\pi)=1619$~MHz~$\mu$m$^3$. We also consider $^{87}$Rb environment atoms, with state choices $\ket{g}=\ket{5s}$, $\ket{e}=\ket{5p}$ and $\ket{r}=\ket{38s}$. The corresponding Rydberg-Rydberg interactions are thus $V_{\alpha\beta}^{(rr)} = C_{6,rr}/(|\alpha-\beta|d)^6$ with dispersion coefficient $C_{6,rr}/(2\pi)=530$~MHz~$\mu$m$^6$, $V^{(pr)}_{n\alpha}=C_{4,pr}/|\bm{R}_n-\bm{R}_\alpha|^4$ with dispersion coefficient  $C_{4,pr}/(2\pi)=-1032$~MHz~$\mu$m$^4$, and $V^{(sr)}_{n\alpha}=C_{6,sr}/|\bm{R}_n-\bm{R}_\alpha|^6$ with $C_{4,pr}/(2\pi)=-87$~MHz~$\mu$m$^6$.

\emph{Distance measures.}---
Here we detail the two distance measures employed in the main text. The first measure is the \emph{fidelity}. The fidelity between two density matrices $\rho_1$ and $\rho_2$ is defined as \cite{nielsen2010}
\begin{equation}\label{eq:overlap_fidelity}
 F(\rho_1,\rho_2) = \Tr\left\{\sqrt{\sqrt{\rho_1}\rho_2\sqrt{\rho_1}}\right\},
\end{equation}
and can be interpreted a generalized measure of the overlap between two quantum states \cite{gilchrist2005}. 

The second measure is related to the \emph{trace distance} via
\begin{equation}\label{eq:trace_fidelity}
 F_D(\rho_1,\rho_2) = 1-D(\rho_1,\rho_2).
\end{equation}
The trace distance $D(\rho_1,\rho_2)$, which can be interpreted as a measure of state distinguishability, is defined as
\begin{equation}
 D(\rho_1,\rho_2) = \frac{1}{2}\Tr\left\{|\rho_1-\rho_2|\right\},
\end{equation}
with $|\rho|=\sqrt{\rho^\dagger \rho}$. 
From the relation $1-F(\rho_1,\rho_2)\leq D(\rho_1,\rho_2)$ \cite{nielsen2010} it follows that $F_D(\rho_1,\rho_2)\leq F(\rho_1,\rho_2)$, which implies that $F_D(\rho_1,\rho_2)$ is a more conservative measure.

\emph{Target-state fidelities and parameters.}---
Table~\ref{tab:fidelities} displays the fidelities and parameters corresponding to the state preparations shown in Figs.~2 to 4 in the main text.
\begin{table*}
\caption{\label{tab:fidelities} Summary of measures $F$ [\eref{eq:overlap_fidelity}] and $F_D$ [\eref{eq:trace_fidelity}] as well as the laser parameters $\Omega_p,\Delta_p,\Omega_c,\Delta_c$. The two measures $F$ and $F_D$ are evaluated at $t=1~\mu$s $(F,F_D)$ and in the steady state $(\tilde{F},\tilde{F}_D)$.}
\begin{ruledtabular}
\begin{tabular}{cccccccc}
N & Target state & $(F,\tilde{F})$ & $(F_D,\tilde{F}_D)$ & $\Omega_p/(2\pi)$ (MHz) & $\Omega_c/(2\pi)$ (MHz) & $\Delta_p/(2\pi)$ (MHz) & $\Delta_c/(2\pi)$ (MHz)\\
2 & $\dyad{\Psi^-}$ & (0.999, 0.999) & (0.992, 0.998) & 7.6 & 95.3 & -75.8 & -44.9\\
2 & $\dyad{\Psi^+}$ & (0.999, 0.999) & (0.997, 0.998) & 8.0 & 89.8 & 65.3 & -7.7\\
2 & $\rho^\mathrm{th}_{T_L}$ & ($>$0.999, $>$0.999) & ($>$0.999, 0.999) & 19.7 & 35.6 & -18.1 & -4.6\\
2 & $\rho^\mathrm{th}_{T_H}$ & ($>$0.999, $>$0.999) & (0.997, 0.997) & 30.0 & 48.9 & -7.3 & -70.4\\
3 & $\rho^\mathrm{th}_{T_L}$ & ($>$0.999, $>$0.999) & (0.994, 0.992) & 9.5 & 34.5 & -7.5 & -71.0\\
3 & $\rho^\mathrm{th}_{T_H}$ & ($>$0.999, $>$0.999) & (0.999, 0.999) & 27.8 & 66.9 & -18.3 & -33.7\\
4 & $\rho^\mathrm{th}_{T_L}$ & (0.996, $>$0.999) & (0.931, 0.990) & 34.6 & 34.9 & -34.5 & -70.6\\
4 & $\rho^\mathrm{th}_{T_H}$ & ($>$0.999, $>$0.999) & (0.994, 0.994) & 35.0 & 43.3 & -1.3 & -45.8\\
\end{tabular}
\end{ruledtabular}
\end{table*}

\emph{Quantum-jump unraveling of the master equation dynamics.}---
A stochastic unraveling of the preparation dynamics provides additional physical insight into the state preparation mechanism. Particularly useful in our case is the quantum jump unraveling, where the occurrence of quantum jumps can be associated with the detection of photons emitted by the environment atoms.

\begin{figure}[tb]
\centering
\includegraphics[width=0.8\columnwidth]{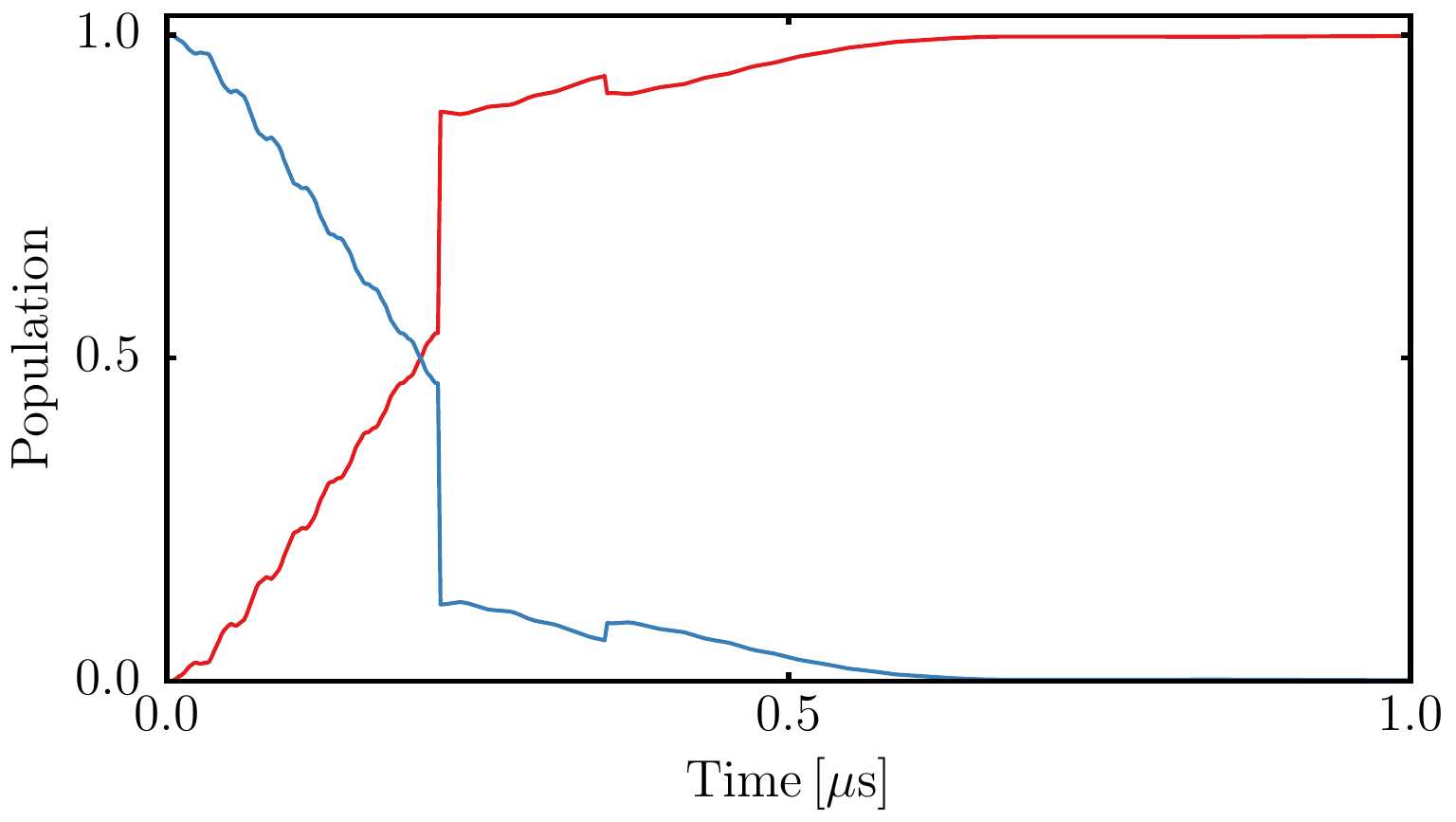}
\caption{Exemplary stochastic trajectory for $\ket{\Psi^-}$ state preparation, given perfect detection of emission from the $\ket{e}$ environment state.
The red (blue) line shows the $\ket*{\Psi^{-(+)}}$ state population.
The jump times are at 0.22~$\mu$s and 0.35~$\mu$s, where sharp features are observed.
The system is initially in state $\ket*{\Psi^{+}}\ket{gg}$.}
\label{fig:mech_corr}
\end{figure}

An exemplary stochastic trajectory is shown in \fref{fig:mech_corr}, corresponding to a quantum jump unraveling of the master equation~(\ref{eq:me}), with distinguishable detection of the decay events from the environment $\ket{e}$ atomic states.
Parameters are chosen for the preparation of the $\ket{\Psi^-}$ state, and are provided in Tab.~I. 
The population in the $\ket{\Psi^-}$ state increases dramatically at the first jump event, due to the population rescaling according to the $\ket{e}$ populations at the appropriate time (see e.g.~\cite{carmichael1991,molmer1993} for a formal treatment of quantum jumps).
The subsequent jump then slightly decreases the target state population, which is quickly restored through coherent dynamics.
Together, the coherent evolution and spontaneous emission from the $\ket{e}$ state drive the system into the $\ket{\Psi^-}$ state.

Note that the detection of the emitted photons can even be used to provide conditional enhancement of the state fidelity~\cite{bentley2014}.

\emph{Dependence of thermal-state quality on system size.}---
To assess the feasibility of thermal-state preparation for larger system sizes, we have evaluated the dependence of thermal-preparation quality $F_D$ with increasing system size. Since due to the long evaluation times numerical optimization of the laser parameters is impractical for system sizes larger than $N\approx 4$, we have adopted a different strategy. That is, we have calculated the thermal-state preparation quality $F_D$ for different sizes, using a fixed laser parameter set.

\begin{figure}[tb]
\centering
\includegraphics[width=0.8\columnwidth]{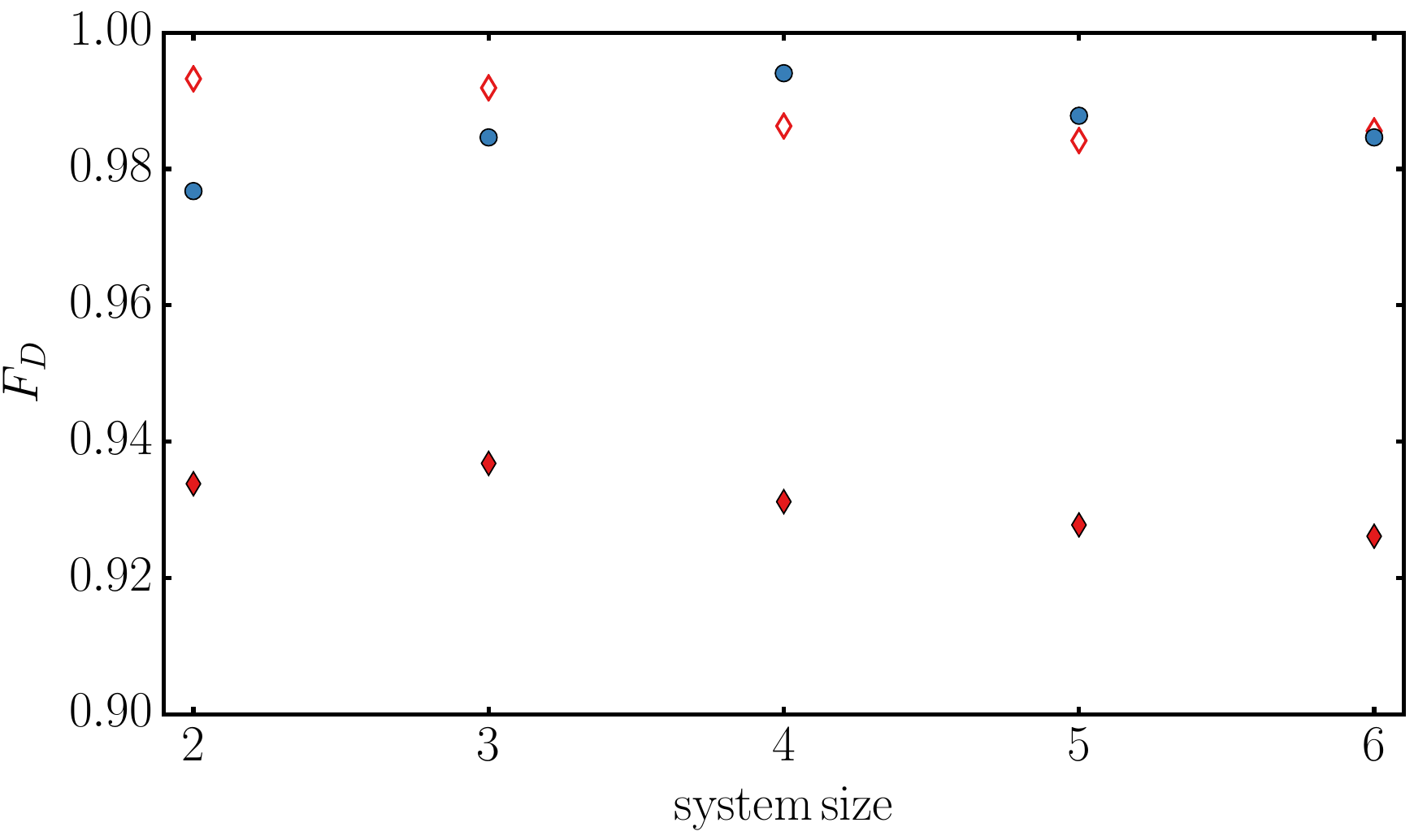}
\caption{Thermal-state preparation quality $F_D$ after $t=1~\mu$s evolution time as a function of system size, both for the low-temperature $T_L$ (filled red diamonds) and high-temperature $T_H$ (blue circles) case. To demonstrate the rapid increase in low-temperature preparation quality with larger preparation times, the low-temperature preparation quality $F_D$ obtained after $t=2~\mu$s is shown with red, empty diamonds.}
\label{fig:scaling}
\end{figure}

In \fref{fig:scaling} we show the dependence of the thermal-state preparation quality $F_D$ on system size for both low- and high-temperature cases ($T_L$ and $T_H$), using the laser parameters listed in Tab.~\ref{tab:fidelities} for $N=4$. Since the dependence of the thermal-state preparation quality on the system size is weak in this case, we expect our preparation scheme to be applicable to even larger system sizes, for which numerical simulation is impeded due to the exponential increase of Hilbert space size with environment atoms. Moreover, the preparation quality in \fref{fig:scaling} does not strictly decrease with increasing system size, which further supports our expectation that state preparation is feasible even for larger system sizes.

\emph{Underlying Mechanism.}---
To obtain an intuitive picture of the preparation mechanism we consider here the smallest possible case: two system atoms and only one environment atom, which is only interacting with one of the system atoms. This allows us to derive the resonance condition for the $\ket{\Psi^+}$ state preparation discussed in the main text, which we do now.

To this end, we consider the total Hamiltonian $\mathcal{H}_\mathrm{tot}$ of system and environment. In the basis ($\ket*{\pi_1,g}$, $\ket*{\pi_1,e}$, $\ket*{\pi_1,r}$, $\ket*{\pi_2,g}$, $\ket*{\pi_2,e}$, $\ket*{\pi_2,r}$), this Hamiltonian reads as
\begin{equation}
\label{eq:Ham_original}
\mathcal{H}_\mathrm{tot} = 
\begin{pmatrix}
0 & \Omega_p/2 & 0 & W & 0 & 0\\
\Omega_p/2 & -\Delta_p  & \Omega_c/2 & 0 & W & 0 \\
0 & \Omega_c/2 & \tilde{V}_{11} & 0 & 0 & W \\
W & 0 & 0 & 0 & \Omega_p/2 & 0 \\
0 & W & 0 & \Omega_p/2 & -\Delta_p & \Omega_c/2 \\
0 & 0 & W & 0 & \Omega_c/2 & \tilde{V}_{21} 
\end{pmatrix},
\end{equation}
with $\tilde{V}_{n\alpha}=\bar{V}_{n\alpha}-\Delta_p-\Delta_c$.

\onecolumngrid

Since we are interested in the preparation of the system eigenstates $\ket{\Psi^\pm}$, we transform \eref{eq:Ham_original} into a new basis ($\ket*{\Psi^-,g}$, $\ket*{\Psi^-,e}$, $\ket*{\Psi^-,r}$, $\ket*{\Psi^+,g}$, $\ket*{\Psi^+,e}$, $\ket*{\Psi^+,r}$),
\begin{equation}
\label{eq:Ham_pm}
\mathcal{H}_\mathrm{tot} = 
\begin{pmatrix}
-W & \Omega_p/2 & 0 & 0 & 0 & 0\\
\Omega_p/2 & -W-\Delta_p & \Omega_c/2 & 0 & 0 & 0 \\
0 & \Omega_c/2 & -W-\Delta_p-\Delta_c+V_\mathrm{sum}/2 & 0 & 0 & V_\mathrm{diff}/2 \\
0 & 0 & 0 & W & \Omega_p/2 & 0 \\
0 & 0 & 0 & \Omega_p/2 & W-\Delta_p & \Omega_c/2 \\
0 & 0 & V_\mathrm{diff}/2 & 0 & \Omega_c/2 & W-\Delta_p -\Delta_c+V_\mathrm{sum}/2
\end{pmatrix},
\end{equation}
where we defined $V_\mathrm{diff}=\bar{V}_{11}-\bar{V}_{21}$ and $V_\mathrm{sum}=\bar{V}_{11}+\bar{V}_{21}$.

To make subsequent analytical treatment particularly simple, we now choose the detuning $\Delta_c$ as $\Delta_c = V_\mathrm{sum}/2$, such that the diagonal elements of the states $\ket{\Psi^\pm,e}$ and $\ket{\Psi^\pm,r}$ coincide.
This allows us to further diagonalize the $\{\ket{e},\ket{r}\}$ subspace of the environment atoms by introducing the (anti-) symmetrized environment states $\ket{\pm} = (\ket{e} \pm \ket{r})/\sqrt{2}$. Accordingly transforming to the basis ($\ket*{\Psi^-,g}$, $\ket*{\Psi^+,g}$, $\ket*{\Psi^-,-}$, $\ket*{\Psi^+,-}$, $\ket*{\Psi^-,+}$, $\ket*{\Psi^+,+}$), \eref{eq:Ham_pm} becomes
\begin{equation}\label{eq:Ham_simple}
\mathcal{H}_\mathrm{tot} = 
\left(\,
\begin{array}{ c c | c c | c c }
- W & 0 & -\Omega_p/\sqrt{8} & 0 & \Omega_p/\sqrt{8} & 0\\
 0 & W  & 0 & -\Omega_p/\sqrt{8} & 0 &  \Omega_p/\sqrt{8} \\\hline
 -\Omega_p/\sqrt{8} & 0 & -W -\Delta_p - \Omega_c/2 & V_\mathrm{diff}/4 & 0 & V_\mathrm{diff}/4 \\
 0 &  -\Omega_p/\sqrt{8} & V_\mathrm{diff}/4 & W -\Delta_p - \Omega_c/2 & V_\mathrm{diff}/4 & 0 \\\hline
 \Omega_p/\sqrt{8} & 0 & 0 & V_\mathrm{diff}/4 & -W -\Delta_p + \Omega_c/2 & V_\mathrm{diff}/4 \\
 0 & \Omega_p/\sqrt{8} & V_\mathrm{diff}/4 & 0 & V_\mathrm{diff}/4 & W -\Delta_p +\Omega_c/2 
\end{array}\,\right).
\end{equation}

Taking a closer look at the diagonal elements of the Hamiltonian (\ref{eq:Ham_simple}), we see that among the $\ket{\Psi^\pm,\pm}$ states, the $\ket{\Psi^-,-}$ state has the lowest energy while the $\ket{\Psi^+,+}$ state has the highest energy. The order of the states $\ket{\Psi^+,-}$ and $\ket{\Psi^-,+}$ depends on the magnitude of $\Omega_c$; for $\Omega_c/2>W$ the energetic order is as indicated in the matrix with the $\ket{\Psi^+,-}$ state being lower in energy than the $\ket{\Psi^-,+}$ state.

Making the $\ket{\Psi^-,g}$ state with energy $-W$ quasi-resonant with the highest-energy state of the $\{\ket{\Psi^\pm,\pm}\}$ manifold is thus a convenient choice to prepare the $\ket{\Psi^+,g}$ state with energy $W$, since this guarantees that the $\ket{\Psi^+,g}$ state is far off-resonant from any state of the $\{\ket{\Psi^\pm,\pm}\}$ manifold and thus does essentially not lose population in the coherent evolution. (The state $\ket{\Psi^+,g}$ also does not lose population due to spontaneous emission since the environment atom is in its ground state.)
  
An approximate condition for this quasi-resonance can be obtained by equating the energy of the highest-energy state with the energy $-W$ of the state $\ket{\Psi^-,g}$. Here we take $\Omega_p/\sqrt{8}\ll W$, such that the states $\ket{\Psi^\pm,g}$ are essentially eigenstates of $\mathcal{H}_\mathrm{tot}$ with energies $\pm W$.

To obtain an approximate expression for the highest-energy state we now diagonalize the two diagonal blocks \begin{equation}
\begin{pmatrix}
-W - \Delta_p \mp \Omega_c/2 & V_\mathrm{diff}/4\\
V_\mathrm{diff}/4 & W - \Delta_p \mp \Omega_c/2
\end{pmatrix}
\end{equation}
using the transformation matrix 
\begin{equation}
S = 
\begin{pmatrix}
\cos(\theta) & -\sin(\theta)\\
\sin(\theta) & \cos(\theta)
\end{pmatrix}
\end{equation}
with $\tan(2\theta)=-V_\mathrm{diff}/(4W)$, which yields the transformed total Hamiltonian $\mathcal{H}_\mathrm{tot}$, given by
\begin{equation}\label{eq:Ham_fulldiag}
\left(\,
\begin{array}{ c c | c c | c c }
- W & 0 & -\Omega_p\cos(\theta)/\sqrt{8} & -\Omega_p\sin(\theta)/\sqrt{8} & \Omega_p\cos(\theta)/\sqrt{8} & \Omega_p\sin(\theta)/\sqrt{8}\\
 0 & W  & \Omega_p\sin(\theta)/\sqrt{8} & -\Omega_p\cos(\theta)/\sqrt{8} & -\Omega_p\sin(\theta)/\sqrt{8} & \Omega_p\cos(\theta)/\sqrt{8} \\\hline
-\Omega_p\cos(\theta)/\sqrt{8} & \Omega_p\sin(\theta)/\sqrt{8} & -\tilde{W} -\Delta_p -\Omega_c/2 & 0 & -(V_\mathrm{diff}/4)^2/\tilde{W} & W V_\mathrm{diff}/(4 \tilde{W}) \\
-\Omega_p\sin(\theta)/\sqrt{8} & -\Omega_p\cos(\theta)/\sqrt{8} & 0 & \tilde{W} -\Delta_p -\Omega_c/2 & V_\mathrm{diff}/(4 \tilde{W}) & (V_\mathrm{diff}/4)^2/\tilde{W} \\\hline
\Omega_p\cos(\theta)/\sqrt{8} & -\Omega_p\sin(\theta)/\sqrt{8} & -(V_\mathrm{diff}/4)^2/\tilde{W} & W V_\mathrm{diff}/(4\tilde{W}) & -\tilde{W} -\Delta_p +\Omega_c/2 & 0 \\
\Omega_p\sin(\theta)/\sqrt{8} & \Omega_p\cos(\theta)/\sqrt{8} & W V_\mathrm{diff}/(4\tilde{W}) & (V_\mathrm{diff}/4)^2/\tilde{W} & 0 & \tilde{W} -\Delta_p +\Omega_c/2
\end{array}\,\right)
\end{equation}
where we defined $\tilde{W}=\sqrt{W^2+(V_\mathrm{diff}/4)^2}$. Taking $\Omega_c/2 > W$ (to ensure that the energetic order is as indicated in $\mathcal{H}_\mathrm{tot}$), a quasi-resonance between the $\ket{\Psi^-,g}$ state and the highest-energy state occurs if $-W = \tilde{W}-\Delta_p+\Omega_c/2$, which can be achieved by choosing $\Delta_p=W+\sqrt{W^2+(V_\mathrm{diff}/4)^2}+\Omega_c/2$. 

In our setup, $V_\mathrm{diff}/4\sim W$, such that $\theta\approx \pi/8$, and \eref{eq:Ham_fulldiag} is approximately given by 
\begin{equation}\label{eq:Ham_fulldiag_simple}
\left(\,
\begin{array}{ c c | c c | c c }
- W & 0 & -0.3\Omega_p & -0.1\Omega_p\ & 0.3 \Omega_p & 0.1 \Omega_p\\
 0 & W  & 0.1\Omega_p & -0.3\Omega_p & -0.1 \Omega_p & 0.3 \Omega_p \\\hline
-0.3\Omega_p & 0.1\Omega_p & -4 W -\Omega_c & 0 & -W/\sqrt{2} & W/\sqrt{2} \\
-0.1 \Omega_p & -0.3 \Omega_p & 0 & -W -\Omega_c & W/\sqrt{2} & W/\sqrt{2} \\\hline
0.3 \Omega_p & -0.1 \Omega_p & -W/\sqrt{2} & W/\sqrt{2} & -4 W & 0 \\
0.1 \Omega_p & 0.3 \Omega_p & W/\sqrt{2} & W/\sqrt{2} & 0 & -W
\end{array}\,\right).
\end{equation}
It is now explicit that the target $\ket{\Psi^+,g}$ state with energy $W$ is far off-resonant from any other state, since its energy difference to any other coupled state is $>2W$, and the corresponding couplings are $\leq 0.3\Omega_p\ll 2W$. The non-target $\ket{\Psi^-,g}$ state, conversely, is coupled to the highest-energy state of the $\{\ket{\Psi^\pm,\pm}\}$ manifold. 
Since the remaining couplings between the states of the $\{\ket{\Psi^\pm,\pm}\}$ manifold are given by $W/\sqrt{2}$, which is much smaller than their energy difference $>2W$, these states are to a good approximation eigenstates of the total Hamiltonian.

For completeness, the Lindblad term in the basis ($\ket*{\Psi^-,g}$, $\ket*{\Psi^+,g}$, $\ket*{\Psi^-,-}$, $\ket*{\Psi^+,-}$, $\ket*{\Psi^-,+}$, $\ket*{\Psi^+,+}$) reads as
\begin{equation}
L = \sqrt{\frac{\Gamma_p}{2}}
\begin{pmatrix}
0 & 0 & -1 & 1 & 0 & 0 \\
0 & 0 & 0 & 0 & -1 & 1 \\
0 & 0 & 0 & 0 & 0 & 0 \\
0 & 0 & 0 & 0 & 0 & 0 \\
0 & 0 & 0 & 0 & 0 & 0 \\
0 & 0 & 0 & 0 & 0 & 0 \\
\end{pmatrix},
\end{equation}
which shows that the states of the $\{\ket{\Psi^\pm,\pm}\}$ manifold decay into the corresponding $\ket{\Psi^\pm,g}$ states. Upon transforming $L$ in the same way as $\mathcal{H}_\mathrm{tot}$ above using the transformation matrix $S$, we obtain 
\begin{equation}
L = \sqrt{\frac{\Gamma_p}{2}}
\begin{pmatrix}
0 & 0 & -\cos(\theta) & \sin(\theta) & \cos(\theta) & -\sin(\theta) \\
0 & 0 & -\sin(\theta) & -\cos(\theta) & \sin(\theta) & \cos(\theta) \\
0 & 0 & 0 & 0 & 0 & 0 \\
0 & 0 & 0 & 0 & 0 & 0 \\
0 & 0 & 0 & 0 & 0 & 0 \\
0 & 0 & 0 & 0 & 0 & 0 \\
\end{pmatrix},
\end{equation}
which shows that the transformed states decay into both $\ket{\Psi^\pm,g}$ states. Accordingly, combining the coherent evolution with the incoherent evolution specified by the Lindblad term, population is transferred out of the non-target state into the target state, which does not lose population due to coherent or incoherent evolution.

Hence, by choosing $\Delta_p=W+\sqrt{W^2+(V_\mathrm{diff}/4)^2}+\Omega_c/2$ we can tune the $\ket{\Psi^-,g}$ state into a quasi-resonance with the highest-energy state of the $\{\ket{\Psi^\pm,\pm}\}$ manifold and thereby prepare the $\ket{\Psi^+}$ state. (We note that $\Delta_p\rightarrow -\Delta_p$ prepares the $\ket{\Psi^-}$ state arguing along the same lines as above.)

\twocolumngrid

\begin{figure}[h]
\centering
\includegraphics[width=0.8\columnwidth]{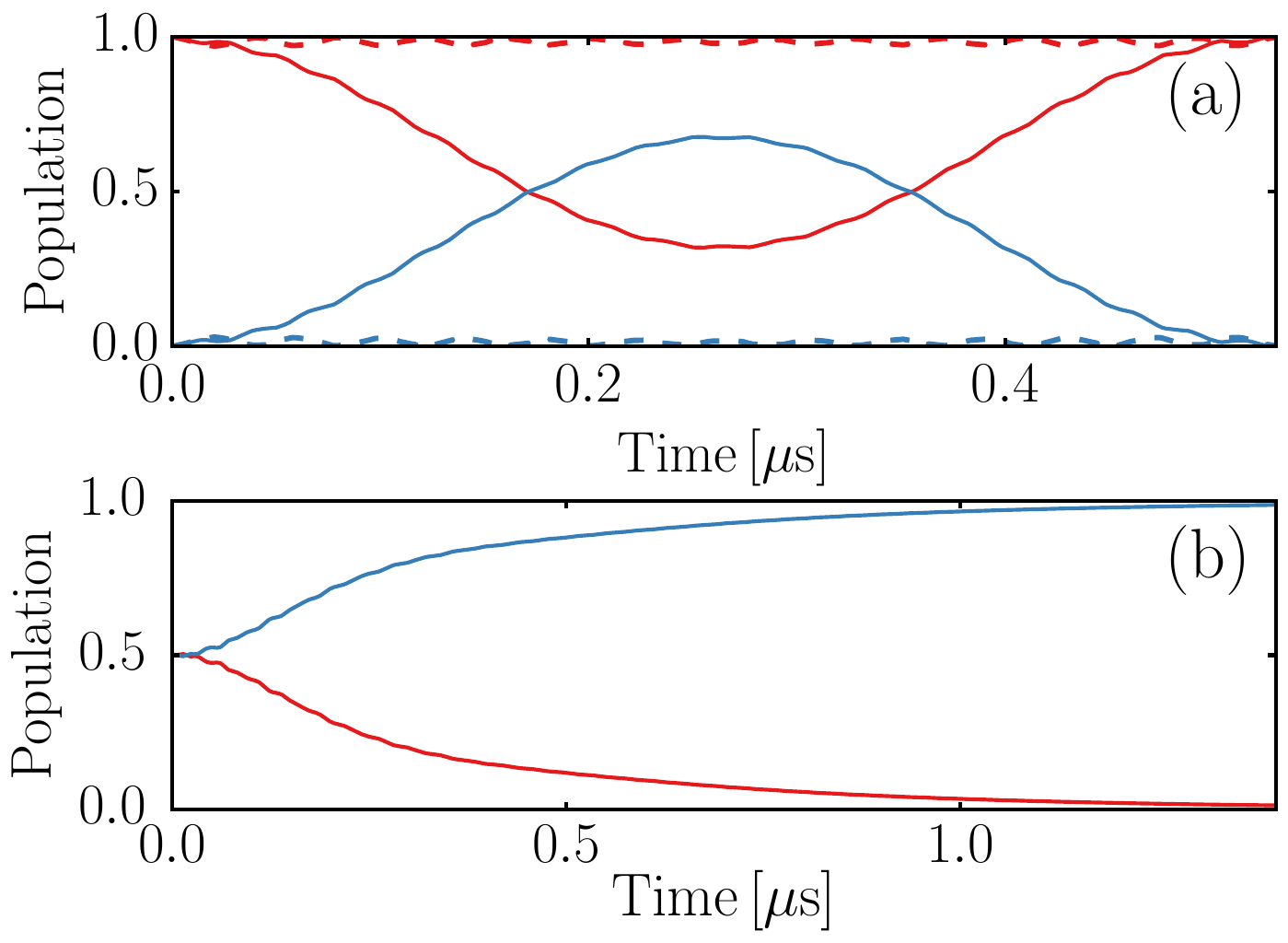}
\caption{(a) Coherent dynamics (neglecting dissipation) corresponding to the preparation of the $\ket*{\Psi^+}$ state using the parameters in the main text. The red (blue) line indicates the population of the non-target $\ket*{\Psi^-,g}$ state (sum of all populations in the $\ket*{\Psi^\pm,\pm}$ states) after initial preparation in $\ket*{\Psi^-,g}$. For comparison, the dashed red (blue) lines show the same for the target $\ket*{\Psi^+,g}$ (sum of all populations in the $\ket*{\Psi^\pm,\pm}$ states) after initial preparation in this state. (b) Corresponding full master equation dynamics including spontaneous emission. The blue (red) line shows the eigenstate populations $\ket{\Psi^+}$ ($\ket{\Psi^-}$). Note the different scaling of the $x$ axis.}
\label{fig:illustration_mechanism}
\end{figure}

To demonstrate that this resonance condition indeed allows us to couple the non-target state to the $\{\ket*{\Psi^\pm,\pm}\}$ manifold while suppressing coherent population transfer from the target state, we show in \fref{fig:illustration_mechanism}(a) the \textit{coherent} population dynamics of non-target state (red) and the $\{\ket*{\Psi^\pm,\pm}\}$ manifold after initial preparation in the non-target state. The parameters here correspond to the $\ket*{\Psi^+}$ state preparation and respect the above resonance condition $\Delta_p$, with $(\Omega_p,\Omega_c,\Delta_p,\Delta_c)/2\pi$: $(7,100,83.4,-32.9)$~MHz. It can be clearly seen that population is coherently transferred out of the non-target state. 
For comparison, the dashed red (blue) lines show the population of the target state ($\{\ket*{\Psi^\pm,\pm}\}$ manifold) after initial preparation in this state. The small magnitude of the oscillations illustrates that population transfer out of the target state is inhibited due to the large detuning of the target state from any other state. Together with the spontaneous emission from the $\{\ket*{\Psi^\pm,\pm}\}$ manifold, this prepares the target state $\ket*{\Psi^+}$, as shown in \fref{fig:illustration_mechanism}(b). The corresponding preparation fidelities are given by ($F$,$\tilde{F}$): (0.98,$>$0.99) and ($F_D$,$\tilde{F}_D$): (0.91,$>$0.99), demonstrating high-fidelity state preparation in the steady state.

We note that in the discussion of the $\ket*{\Psi^+}$ state-preparation mechanism above we imposed constraints on the laser parameters $\Omega_p$, $\Omega_c$ and $\Delta_c$. These constraints are not vital for high-fidelity state preparation, but were chosen for the ease of explanation. 
When adding more system and/or environment atoms, the complexity of the corresponding Hamiltonian rapidly increases, and analytical treatment becomes cumbersome. Moreover, for thermal-state preparation or the preparation on a fast timescale, the constraints on $\Omega_p$, $\Omega_c$ and $\Delta_c$ outlined above are too restrictive when aiming at obtaining high preparation fidelities or designing special environment couplings and dynamics. The additional degrees of freedom in choosing laser parameters in this case render analytical treatment even more cumbersome. In the Letter we therefore use numerical optimization to obtain suitable laser parameter sets.

\end{document}